\documentclass{iopjournal}

\usepackage{physics}
\usepackage{tabularx}
\usepackage{array}
\usepackage{xcolor}
\usepackage{pdfpages}
\newcolumntype{L}[1]{>{\raggedright\arraybackslash}m{#1}}
\newcolumntype{C}[1]{>{\centering\arraybackslash}m{#1}}
\begin{document}

\articletype{Preprint} 

\title{Permeability heterogeneity and bulk linear elasticity of displaced clay suspensions determine interfacial pattern morphologies in Hele-Shaw experiments}

\author{Vaibhav Raj Singh Parmar$^1$\orcid{0000-0003-0895-6106} and Ranjini Bandyopadhyay$^{1,*}$\orcid{0000-0002-4290-6404}}

\affil{$^1$Soft Condensed Matter Group, Raman Research Institute, C. V. Raman Avenue, Sadashivanagar, Bangalore 560 080, India}

\affil{$^*$Author to whom any correspondence should be addressed.}

\email{ranjini@rri.res.in}


\begin{abstract}
When a less viscous Newtonian fluid displaces an aging aqueous clay suspension in a confined space, a rich array of interfacial patterns emerges due to a predominantly viscous instability. In the present work, we controlled the mechanical properties of clay suspensions by incorporating additives and studied the interfacial instabilities that resulted when these suspensions were radially displaced by water in a Hele-Shaw cell. When the elasticity of clay was low, the interfacial dynamics exhibited features of nonlinear viscous fingering in heterogeneous media. By tuning the nature and content of additives that delay clay aging, we uncovered two novel propagation mechanisms: pattern growth \textit{via} skewering and zig-zag finger propagation. These patterns have hitherto never been observed in experiments with colloidal systems. For moderate clay elasticities, we demonstrate here that shear-thinning-induced flow anisotropy leads to the formation of dendrites with dominant side branches. As clay elasticity increases due to the incorporation of salts, the energy required to create fractures becomes smaller than that for system-wide yielding. This scenario is characterized by the emergence of viscoelastic fractures. Our work demonstrates that incorporating additives is an effective strategy to manipulate the onset and growth of interfacial instabilities during the confined displacement of clay by miscible Newtonian fluids.
\end{abstract}

\section{\label{sec:intro}Introduction}
Fluid flow in porous media is ubiquitous in many industrial and natural processes~\cite{doi:10.1073/pnas.1901619116,doi:10.1098/rsta.2016.0155}. When a less viscous fluid displaces a more viscous fluid in a quasi-two-dimensional geometry, the interface becomes unstable and intricate fingering patterns form due to a viscous fingering instability~\cite{saffman1958penetration,doi:10.1146/annurev.fl.19.010187.001415,doi:10.1126/science.243.4895.1150,bischofberger2014fingering}. Understanding such instabilities is important in various fields, such as in enhanced oil recovery processes~\cite{kargozarfard2019viscous}, carbon dioxide sequestration~\cite{gooya2019unstable} and enhanced mixing at low Reynolds numbers~\cite{Fluid_mixing}. In such displacement experiments, finger propagation, fingertip splitting and pattern morphology depend on several factors, such as viscosity contrast and interfacial tension between the fluid pair~\cite{saffman1958penetration, bischofberger2014fingering, 10.1063/5.0030152}, the injection flow rate or injection pressure of the less viscous displacing fluid~\cite{PhysRevLett.67.2005}, and the gap of the Hele-Shaw (HS) cell~\cite{nand2022effect}. The interfacial patterns that form during miscible displacement experiments (characterized by interfacial tension between the fluid pair tending to zero) are more complex than those formed by immiscible displacement~\cite{chen1987radial}. The emergence of complex patterns can also be driven by local heterogeneities of the displaced medium. In sharp contrast to reports on interfacial pattern formation in homogeneous media, numerical simulations of viscous fingering in heterogeneous media~\cite{10.1063/1.475259,https://doi.org/10.1002/ceat.202300376} have uncovered a novel channeling regime, in which a finger advances by preferentially choosing high-permeability paths.

The complexity of interfacial patterns increases dramatically when either or both fluids are non-Newtonian~\cite{Parmar_2024}. The shear-thinning rheology~\cite{10.1063/1.870303,PhysRevLett.80.1433}, non-zero yield stress~\cite{ESLAMI201925,PhysRevLett.85.314} and elasticity~\cite{MORA201230,PhysRevLett.67.2009} of a non-Newtonian fluid in a displacement experiment can significantly alter interfacial pattern morphologies~\cite{Parmar_2024,PALAK2023130926,PALAK2021127405,ozturk2020flow,PALAK2022100047}. When a shear-thinning fluid is displaced~\cite{PhysRevLett.80.1433}, the velocity of the propagating finger is maximum at the fingertip. Therefore, the resistance to finger growth is minimal in the forward direction, which leads to the formation of narrow fingers and delays the onset of tip-splitting. Given the anisotropic flow conditions, the lower viscosity in the flow direction compared to that in the lateral direction results in the formation of side branches and dendritic patterns~\cite{PhysRevLett.80.1433,PALAK2022100047}. When an yield stress fluid is displaced, fractures develop due to an elastic response below their yielding threshold, while viscous fingers appear above the yielding point~\cite{ESLAMI201925,PhysRevLett.85.314}.

Here, we generate an array of interfacial patterns by displacing aging aqueous Laponite\textsuperscript{\textregistered} clay suspensions by miscible water in a confined, radial Hele-Shaw geometry. We control the physical aging behavior of the clay, a model soft glassy material, by incorporating chaotropic and kosmotropic salts~\cite{D1SM00987G} in the suspension medium. Laponite clay is composed of disk-shaped nanoparticles with an average diameter of $25$-$30$ nm and a thickness of $1$ nm. In an aqueous medium, these clay particles carry negative charges on their faces and positive charges on the rims below their isoelectric point at pH$<11$~\cite{AU201565,PhysRevE.66.021401,C0SM00590H}. Debye screening between Laponite particles leads to short-range electrostatic repulsions between the edges and faces of neighboring particles, plus edge-face electrostatic attractions. The mechanical properties of aqueous clay suspensions lying in the concentration range between 1.0\% w/v and 4.0\% w/v have been reported to evolve over time in a physical aging process that is driven by the gradual self-assembly of the clay particles into system-spanning networks of interconnected particles ~\cite{PhysRevE.64.021510,B916342E,C0SM00590H}. It has been shown that the incorporation of different additives, such as salts~\cite{doi:10.1021/acs.langmuir.5b00291,D1SM00987G}, acids~\cite{THRITHAMARARANGANATHAN2017304} and non-dissociative molecules~\cite{D1SM00987G} in the suspension medium can accelerate or delay the aging dynamics of clay. By engineering the initial age of the displaced clay 
and by changing the injection rate of the displacing fluid~\cite{van1986fractal,PhysRevLett.89.234501}, the rich time-dependent (aging) rheological properties of clay can be exploited to generate an array of exotic interfacial patterns in displacement experiments. Indeed, we had reported in earlier studies that miscible displacements of soft glassy clay suspensions of increasingly higher ages resulted in interfacial patterns that transformed from dense viscous and dendritic to viscoelastic fractures~\cite{PALAK2022100047, PALAK2023100084}. It should therefore be possible to generate an even broader range of novel interfacial patterns during miscible Hele-Shaw displacements of aqueous clay \textit{via} the controlled incorporation of additives to the aqueous medium in which the clay was prepared.

\renewcommand{\arraystretch}{1.3}
\begin{table*}[!b]
\small
\centering
\begin{tabularx}{1.025\textwidth}{|L{2.9cm}|L{2.4cm}|C{2.4cm}|L{3.2cm}|L{2.6cm}|}
\hline
\textbf{Additive name} & \textbf{Additive nature} & \textbf{Concentration used} & \textbf{Mechanism by which clay aging is modified} & \textbf{Influence on clay aging}  \\
\hline
Tetrasodium pyrophosphate (TSPP) & Dissociative and kosmotropic & 1--2 mM & Inhibits edge-face bonding by adsorbing on clay rims & Delays/inhibits clay aging \\
\hline
Dimethylformamide (DMF) & Nondissociative and chaotropic & 130--260 mM & Disrupts medium structure & Delays clay aging \\ 
\hline
Sodium chloride (NaCl) & Dissociative and kosmotropic & 1--2 mM & Screens electrostatic repulsion & Accelerates clay aging \\ 
\hline
Potassium chloride (KCl) & Dissociative and chaotropic & 1--2 mM & Screens electrostatic repulsion & Accelerates clay aging \\ 
\hline
Glucose & Nondissociative and kosmotropic & 130--260 mM & Promotes medium structure & Accelerates clay aging \\ 
\hline
\end{tabularx}
\caption{\label{fig:table}List of additives used in this article, their nature, concentrations in the clay samples used here, the mechanisms by which clay aging is modified and the influence of each additive on clay aging. }
\end{table*}
The additives that were chosen to be incorporated in the suspension medium in the present work are known to either delay or accelerate clay aging (\textit{i.e.}, they exhibit chaotropic and kosmotropic actions respectively). When additives that delay clay aging, such as dimethylformamide (DMF) and tetrasodium pyrophosphate (TSPP)~\cite{D1SM00987G,MONGONDRY2004191}, were introduced in the medium while preparing aqueous clay, the growth of interfacial patterns due to the displacement of clay by water was primarily dominated by tip-splitting events, with the finger propagation profile seen to be highly sensitive to the nature of the additive used. We also observed exotic nonlinear viscous fingering mechanisms like skewering, dense-branching, coalescence, trailing lobe detachment, chanelling and zig-zag finger propagation. In contrast, when additives like sodium chloride (NaCl), potassium chloride (KCl) and glucose, known to enhance clay aging and therefore sample elasticity, were added to the suspension medium, the interfacial pattern morphologies were dominated by side branches and fractures.  The patterns in these experiments were sensitive only to the concentration of the additive rather than to its exact chemical composition. 

We characterized the distinct interfacial pattern morphologies by computing the time-dependent areal ratios, $A_p/A$~\cite{PALAK2022100047, PALAK2023100084}, a measure of pattern compactness in 2D, the instantaneous velocity of the longest fingertip, $U$, and the total number of fingertips, $N$. 
In a significant advance to existing studies, we show here that an even richer array of patterns with exotic morphological features, including some that were previously unseen in colloidal systems, may be generated in miscible Hele-Shaw displacement experiments involving clay and water. We achieved this by changing the heterogeneity, permeability, and rheological behavior of the displaced clay suspension \textit{via} the incorporation of chaotropic and kosmotropic additives in the suspension medium.

\section{\label{sec:mnm}Material and Methods}
\begin{figure}[!b]
  \centering
  \includegraphics[width=0.49\textwidth]{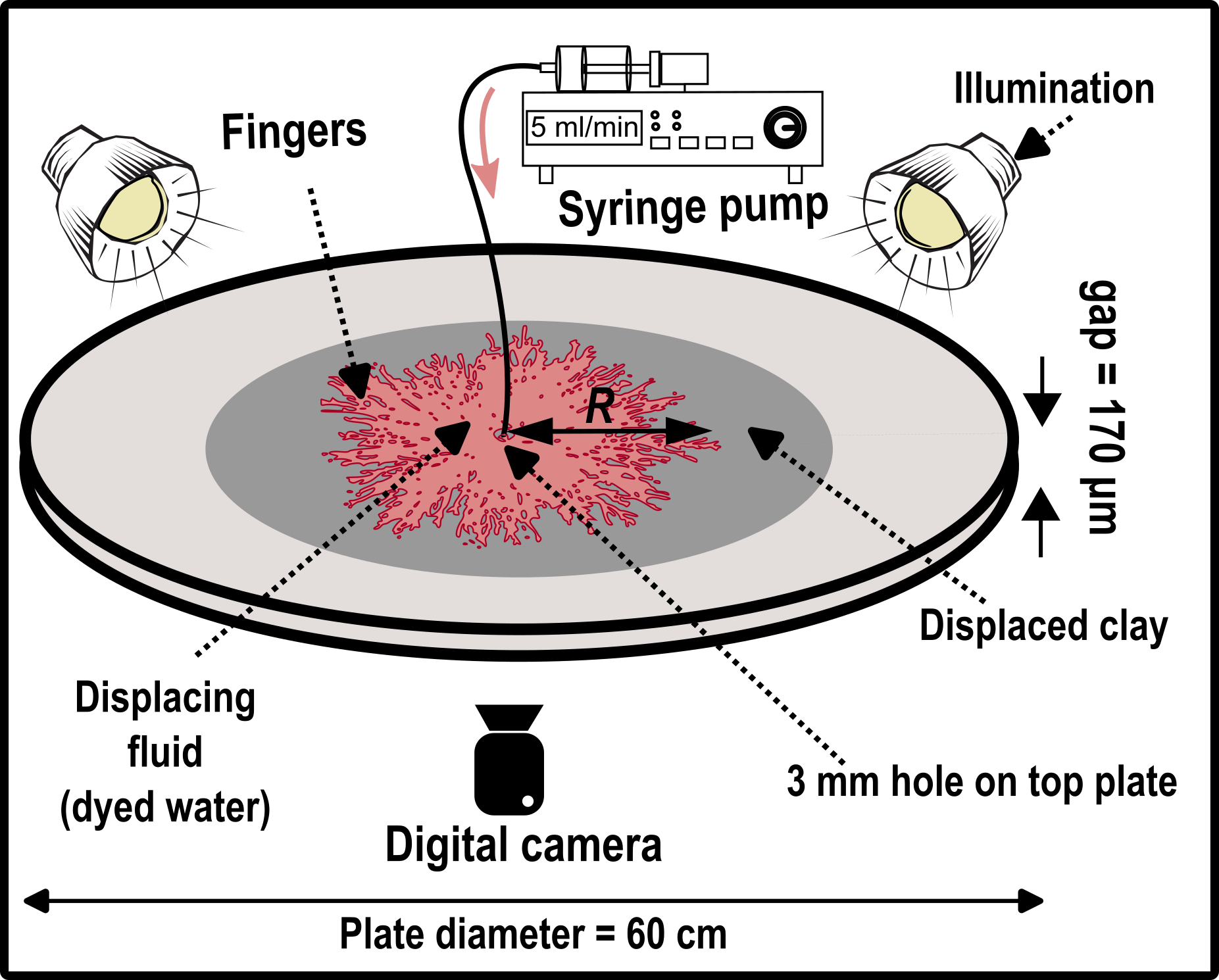}
  \caption{Schematic illustration of the radial Hele-Shaw cell geometry with plate separation $b$ = $170$ $\mu$m used in displacement experiments. Here, the gray region represents the displaced clay, while the pink region represents the displacing fluid (water). A small amount of dye was added to the displacing fluid to visualize the interface clearly. The length of the longest propagating finger is denoted by $R$.}
  \label{fig:1}
\end{figure}
\subsection{\label{sec:level2} Preparation of aqueous Laponite clay}
Moisture from Laponite\textsuperscript{\textregistered} XLG powder (purchased from BYK Additives Inc.) was removed by baking it at 120$^\circ$C for $18$ - $20$ hours in an oven. Aqueous media were prepared by mixing desired concentrations of additives with de-ionized water (Millipore Corp., resistivity $18.2$ M$\Omega$-cm). NaCl (LABORT Fine Chem Pvt. Ltd), KCl (Sigma-Aldrich), glucose (Sigma-Aldrich), dimethylformamide (DMF, SDFCL Fine Chem Pvt. Ltd) and tetrasodium pyrophosphate (TSPP, E. Merck (India) Ltd) were used as received without further purification. Laponite clay suspensions were freshly prepared by weighing predetermined amounts of dried Laponite powder, which were gradually added to an aqueous medium while stirring vigorously to prepare pure clay samples of concentration $3.0$\% w/v. Solutions with additives at predetermined concentrations were prepared to which clay was added to produce new batches of $3.0$\% clay suspensions. The rheology of the suspensions was strongly dependent on the nature and content of the additive. The presence of NaCl, KCl, and glucose in the suspension medium promotes the formation of system-spanning microstructures and accelerates the physical aging of clay~\cite{D1SM00987G}. In contrast, DMF and TSPP delay or inhibit clay microstructure formation~\cite{D1SM00987G,MONGONDRY2004191} and delay the physical aging process. A list summarizing the nature of these additives and their influence on clay aging is shown in Table~\ref{fig:table}. After $45$ minutes of vigorous stirring, the transparent clay samples were filtered through syringe filters of pore size $0.45$ $\mu$m (Millex\textsuperscript{\textregistered}, Sigma Aldrich) and immediately loaded into experimental geometries for rheological and displacement experiments. The pH values of all the clay samples were measured using an Eutech 2700 pH meter. Displacement experiments were performed in radial Hele-Shaw cells~\cite{HELE-SHAW1898}, wherein aqueous clay suspensions, with and without additives, were displaced by a miscible fluid. Deionized water from Millipore Corp., used as the displacing Newtonian fluid in our experiments, was dyed by dissolving a small amount of Rhodamine B (Sigma-Aldrich) to enhance interfacial contrast in displacement experiments.

\subsection{\label{sec:hs} Radial Hele-Shaw cell used in displacement experiments}
Displacement experiments were performed in a quasi-two-dimensional radial Hele-Shaw (HS) cell consisting of two horizontally arranged parallel glass plates, each of radius 30 cm and thickness 1 cm, as shown in Fig.~\ref{fig:1}. The gap between the two circular glass plates of the HS cell was maintained at $170$ $\mu$m with Teflon spacers. A freshly prepared clay sample was filled in the HS cell through the $3$ mm hole at the center of the top plate. The time at which suspension loading was completed was noted down as aging time $t_w = 0$ h. The clay sample was left undisturbed inside the HS cell and allowed to age spontaneously for $t_w = 5$ h. At this instant, the displacing fluid (dyed water) was injected through the same central hole at a constant injection flow rate $q$ = $5$ ml/min using a syringe pump (NE-8000, New Era Pump Systems, USA). The evolution of interfacial patterns was recorded at $29.97$ frames per second using a Nikon D5200 DSLR camera. The image sequences for the observed interfacial patterns were analyzed using the MATLAB@2019 image processing toolbox. All experiments were performed at room temperature.
\begin{figure}[!b]
		\includegraphics[width=0.99\textwidth]{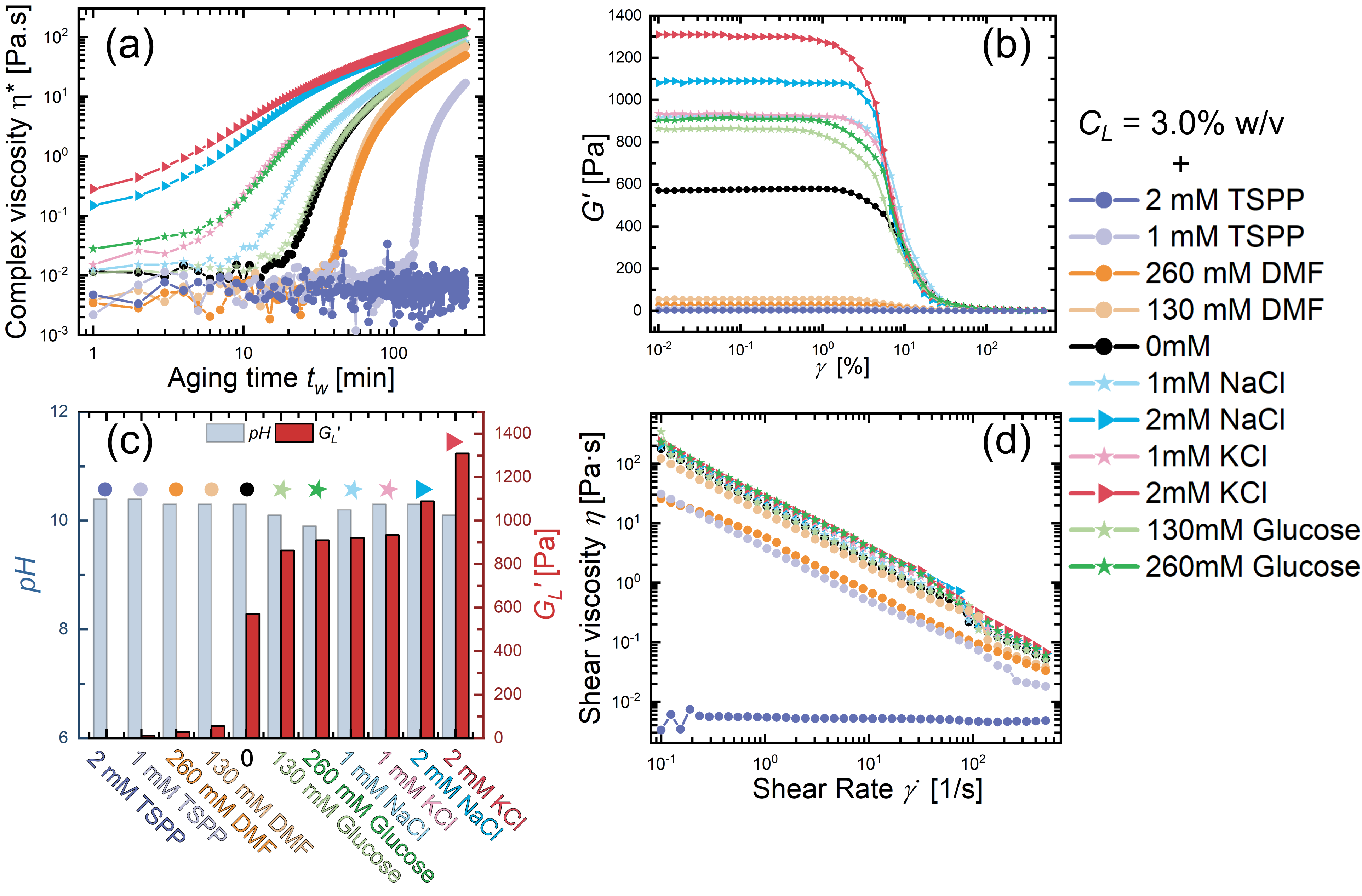}
		\centering
		\caption{\label{fig:2} (a) Complex viscosity $\eta^*$ versus aging time $t_w$ for clay samples at $3.0$\% w/v clay concentration, prepared in aqueous solutions containing various chaotropic and kosmotropic additives. (b) The storage moduli $G^\prime$ versus shear strain $\gamma$, obtained from amplitude sweep oscillatory rheological measurements for the same samples at $t_w = 5$ h. (c) pH (at $t_w = 0$ h) and plateau elastic modulus ${G_L}^\prime$ (at $t_w = 5$ h) for all the clay samples. (d) Steady-state flow curves of shear viscosity $\eta$ as a function of shear rate $\dot{\gamma}$ for all the clay samples at $t_w = 5$ h.
  }
  \end{figure}
\subsection{\label{sec:rheo} Rheological experiments to measure elastic and viscous moduli of aqueous clay}
An Anton Paar MCR 702 rheometer was used to perform rheological measurements. A freshly prepared clay suspension was loaded in the measuring geometry and shear melted at a high shear rate $\dot{\gamma}$ = $500$ s$^{-1}$ for $2$ minutes to ensure identical initial conditions in all rheological experiments. The time at which shear melting was stopped was noted as aging time $t_w$ = $0$ h. A parallel plate (PP50) measuring geometry was used to perform time sweep and flow curve measurements, while a double gap geometry (DG26.7) was used for amplitude sweep measurements. 
Time sweep measurements were initiated immediately after shear melting was stopped. For the amplitude sweep and flow curve experiments, in contrast, the clay sample was left undisturbed in the measuring geometry and allowed to age to $t_w$ = $5$ h. The suspension temperature was maintained at $22$$^\circ$C throughout the experimental duration using a Peltier device and a water circulation system (Viscotherm VT2). Silicone oil of viscosity $5$ cSt was used as a solvent trap to prevent water evaporation during the measurements.
\section{\label{sec:rnd}Results and discussion}
 \subsection{\label{sec:rheo2}Controlling the aging behavior of Laponite clay by incorporating additives}
We measured the rheological responses of the pure clay sample and the clay samples with additives, \textit{viz.}, dimethylformamide (DMF), tetrasodium pyrophosphate (TSPP), sodium chloride (NaCl), potassium chloride (KCl) and glucose, incorporated in the aqueous suspension medium. Figure~\ref{fig:2}(a) shows the time evolution of the complex viscosity, defined as $\eta^* = \sqrt{{G^\prime}^2 + {G^{\prime\prime}}^2}/\omega$, for each clay sample studied in this work. In the expression for $\eta^*$, $G^\prime$ and $G^{\prime\prime}$ are the elastic and viscous moduli of the samples, respectively, measured using oscillatory time-sweep experiments at an angular frequency, $\omega$, of $6$ rad/s and shear strain amplitude, $\gamma$, of $0.1$\%. Figure~\ref{fig:2}(b) presents the storage modulus $G^\prime$ as a function of strain amplitude for all the samples at an aging time of $t_w = 5$ h. The corresponding loss moduli $G^{\prime\prime}$ obtained from these amplitude-sweep experiments are provided in Supplementary Fig.~S1. The linear elastic moduli $G_L^\prime$, which represent the elastic response of the sample and are extracted from the linear rheological data as described in Fig.~S2 of the Supplementary Information, are shown in Fig.~\ref{fig:2}(c). Fig.~\ref{fig:2}(c) also displays the pH values of all the suspensions. We note that the measured pH values all lie between $9.9$ and $10.4$. Since the Laponite isoelectric point is $11$, this observation indicates that the Laponite disks in all the clay samples are negatively charged while their rims are positively charged~\cite{AU201565}. It is therefore expected that the clay particles are self-assembled in the aqueous medium \textit{via} overlapping coins or house-of-card particle associations. Finally, Fig.~\ref{fig:2}(d) shows the steady-state shear viscosity $\eta$ as a function of shear rate $\dot{\gamma}$ for all the suspensions, estimated from the flow curves measured in rotational rheology experiments.

The addition of both DMF and TSPP hinders the development of microstructures and delays physical aging. DMF disrupts the hydrogen-bonding network in water and limits tactoid swelling, which accelerates particle dynamics and delays suspension aging~\cite{D1SM00987G}. TSPP, on the other hand, neutralizes the positive rim charges on the clay particles \textit{via} the attachment of pyrophosphate ($P_2O_7^{4-}$) groups. This inhibits edge-face associations and delays network formation~\cite{PhysRevE.66.021401,MONGONDRY2004191}. Therefore, even though the mechanisms are quite distinct, DMF and TSPP both have a similar effect on clay aging. The extremely low steady-shear viscosity and the absence of shear-thinning behavior in the $2$ mM TSPP sample (blue circles in Fig.\ref{fig:2}(d)) both indicate a lack of microstructural development. We note that the low elastic moduli observed in clay at $t_w = 5$ h with DMF and TSPP added to the medium (Fig.\ref{fig:2}(b)) are consistent with their delayed gelation behavior.

In contrast, the addition of NaCl, KCl and glucose significantly accelerates the aging process by enhancing the buildup of self-assembled suspension microstructures~\cite{doi:10.1021/acs.langmuir.5b00291,D1SM00987G}. Monovalent salts like NaCl and KCl reduce the Debye screening length, which promotes edge-face interactions between clay particles and eventually facilitates the formation of space-spanning networks of self-assembled clay particles~\cite{doi:10.1021/acs.langmuir.5b00291,D1SM00987G,doi:10.1021/acs.langmuir.8b01830}. Glucose accelerates clay aging \textit{via} a non-DLVO mechanism by forming hydrogen bonds with water molecules and effectively reducing the amount of free water available to hydrate the clay~\cite{D1SM00987G,Shoaib2023}. As a consequence, clay samples with glucose exhibit higher complex viscosities (green stars in Fig.\ref{fig:2}(a)), higher storage moduli (Fig.~\ref{fig:2}(b)) and exhibit shear-thinning behavior (Fig.~\ref{fig:2}(d)). The rheological signatures in Figs.~\ref{fig:2}(a–c) are all consistent with the hypothesis that NaCl, KCl and glucose speed up clay aging by accelerating microstructural buildup. In contrast, the rheological signatures for DMF and TSPP are indicative of delayed suspension aging. Finally, we note that while clay elasticities are vastly different due to the incorporation of additives, they do not change appreciably during our displacement experiments, which last between $3$ and $40$ seconds.
 \begin{figure*}[!t]
		\includegraphics[width=1.0\textwidth]{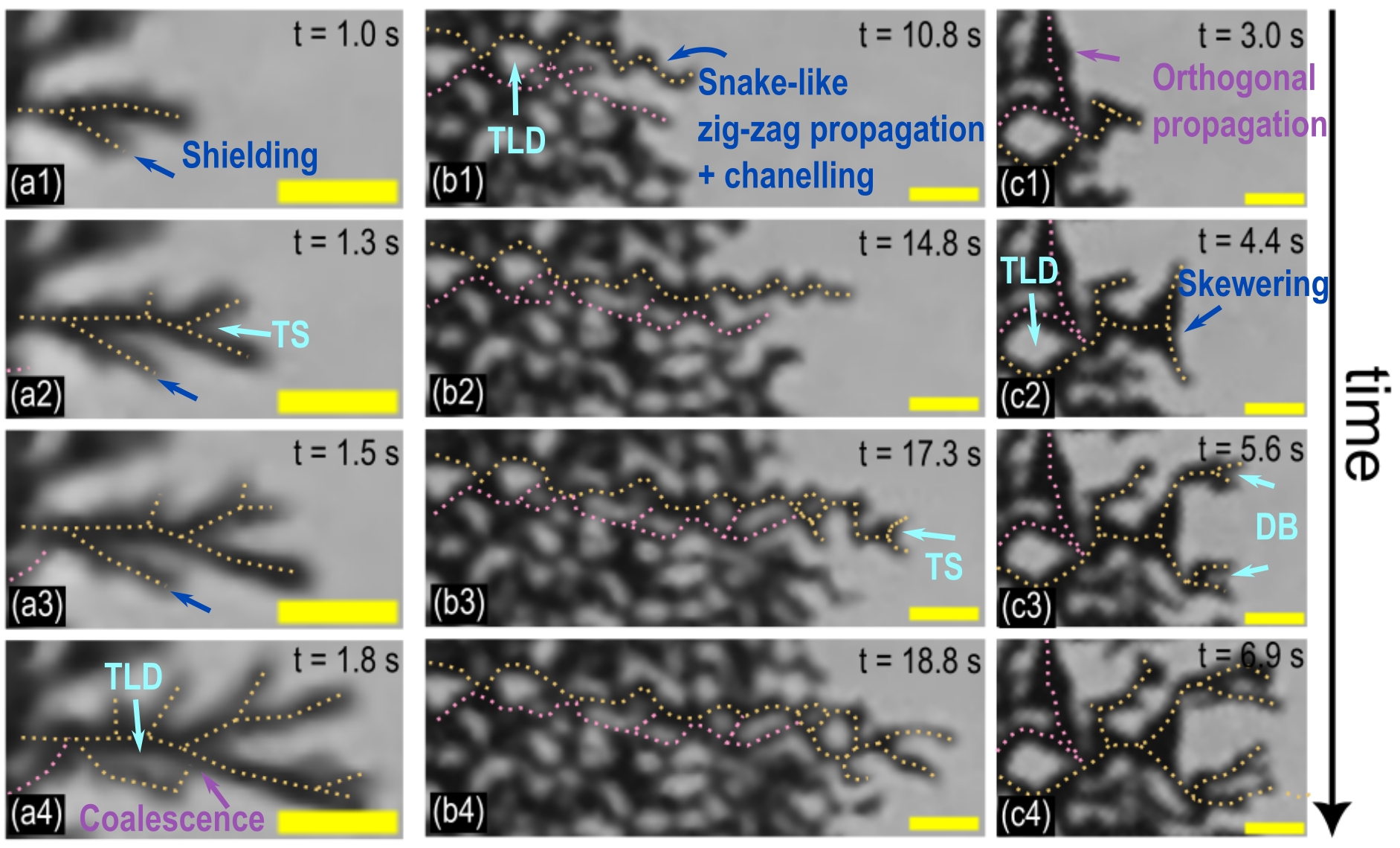}
		\centering
		\caption{\label{fig:3} Cropped grayscale images displaying the temporal propagation profiles of fingers formed during the miscible displacements of $3.0$\% w/v clay at $t_w = 5$ h. The clay samples in these experiments were prepared \textbf{(a1-a4)} without additives, with \textbf{(b1-b4)} $2$ mM TSPP and \textbf{(c1-c4)} $260$ mM DMF. Dotted lines have been drawn to highlight the propagation mechanisms of the interfacial fingers. The scale bar in each image is $0.5$ cm. The abbreviations used are as follows: TS: Tip-splitting, TLD: Trailing lobe detachment and DB: Dense branching.
  }
  \end{figure*}
\subsection{Finger propagation profiles during miscible displacement of pure clay}
  Figures~\ref{fig:3}(a1-a4) show the sequential evolutions of finger propagation profiles, with yellow and pink dashed lines highlighting the propagation of two neighboring fingers, when a clay sample of concentration, $C_L=$ 3\% w/v, was displaced by water. The movie from which these snapshots were extracted is provided as Supplementary Movie~1. The interface between the water and the clay is poorly defined due to fluid mixing. We observed finger shielding (Figs.~\ref{fig:3}(a1-a3)), with one advancing finger retarding the growth of a neighboring finger. Additionally, as displayed in Fig~\ref{fig:3}(a2), the fingers underwent consecutive tip-splitting (TS) events, with the fastest propagating fingers splitting at their tips to form two distinct fingers. We also identified finger coalescence events (Fig.~\ref{fig:3}(a4)), wherein the tip of the trailing finger bent and merged with a neighboring finger. Multiple coalescence events eventually led to the detachment of small portions of the displaced clay from the main body in a mechanism known as trailing lobe detachment (TLD), also marked in Fig.~\ref{fig:3}(a4). Coalescence and the formation of trailing lobes are intricately linked to each other. It is well known that a propagating finger that is shielded by a neighboring finger witnesses a loss of momentum, merges with the latter and results in the formation of trailing lobes that eventually detach from the main body~\cite{https://doi.org/10.1002/fld.803,10.1063/1.858728,Ghesmat2008}. We note from our experiments that these trailing lobes, which comprise the displaced clay, diffuse into the displacing fluid over time. 
  
  Non-linear finger interactions such as shielding, tip-splitting, coalescence, and trailing lobe detachment have been widely reported in simulations of Newtonian fluid pairs characterised by a high mobility ratio~\cite{https://doi.org/10.1002/fld.803}, experiments involving the displacement of glycerol by water~\cite{https://doi.org/10.1002/cjce.5450840109} and during the displacements of non-Newtonian fluids in confined geometries~\cite{Azaiez_2004}. We note from our experiments that when an aqueous clay without any salts/additives is displaced by water, the viscosity contrast between clay and water results in the growth of interfacial patterns that are driven by the Saffman-Taylor instability~\cite{saffman1958penetration}.

\subsection{Finger propagation profiles during miscible displacements of clay samples prepared with TSPP and DMF}
As already outlined in Table~\ref{fig:table}, the addition of TSPP and DMF in the aqueous medium while preparing clay suspensions delays clay aging by retarding the self-assembly of clay particles into networks. Figures~\ref{fig:3}(b1-b4, c1-c4) illustrate the sequential evolutions of finger propagation profiles for clay samples of concentration $C_L =$ 3\% w/v in the presence of $2$ mM TSPP and $260$ mM DMF, respectively. The movies from which these snapshots were extracted are provided as Supplementary Movies~2 and 3, respectively. Comparison with Fig.~\ref{fig:2}(c) shows that the displaced clay in these experiments was characterized by low plateau elastic moduli ranging between $2.0 \leq {G_L}^\prime \leq 55.4$ Pa. 

\begin{figure}[!t]
		\includegraphics[width=0.99\textwidth]{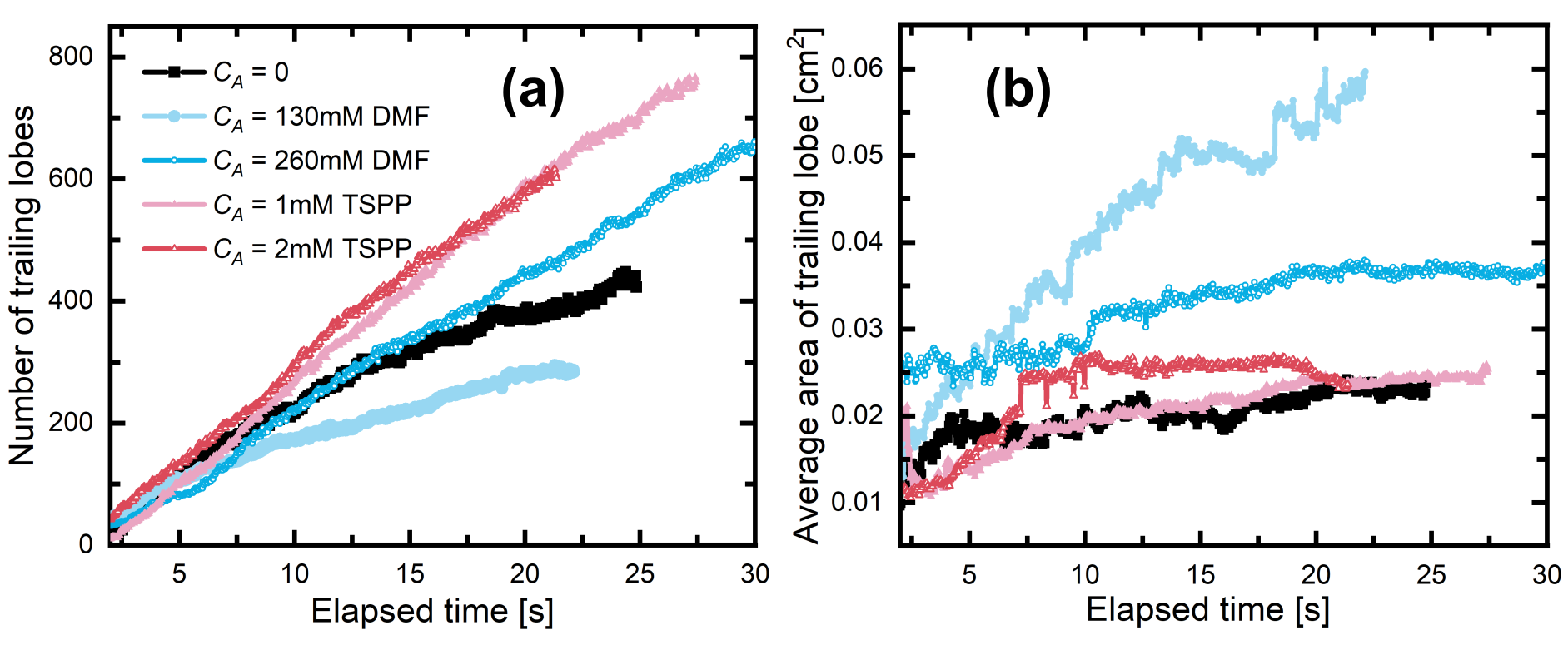}
		\centering
		\caption{\label{fig:5} The images in Figs.~\ref{fig:3}(a-c) were analyzed to estimate the temporal evolution of the (a) number of trailing lobes and (b) average areas of trailing lobes at the clay-water interface when 3.0\% w/v clay samples with TSPP and DMF in the suspension medium were displaced by water.
  }
  \end{figure}
  Images of interfacial instabilities with clay samples containing TSPP and DMF are provided in Fig.~S3 of the Supplementary Information. The global morphology of the interfacial patterns observed in the presence of TSPP and DMF can be categorised as a dense viscous pattern (DVP). Figs.~\ref{fig:3}(b1-b4) display magnified views of the propagating fingers with 2 mM TSPP and reveal that the fingers propagate in a snake-like zig-zag manner, as shown by pink dashed lines, with relatively fewer tip-splitting events. The finger growth profile, in this case, is reminiscent of channeling~\cite{10.1063/1.475259,https://doi.org/10.1002/ceat.202300376}, wherein interfacial fingers propagate through preferential flow pathways through the displaced suspension. Chanelling has been reported for displacements in heterogeneous porous media for which pathways for preferential flows are synonymous with regions of high permeability~\cite{10.1063/1.475259,https://doi.org/10.1002/ceat.202300376}.

In contrast, the images acquired in the experiments with DMF (Figs.~\ref{fig:3}(c1-c4)) reveal a very different mode of finger propagation. In these displacement experiments, we observed that a growing finger suddenly stopped propagating radially outward. Next, the tip widened (Fig.~\ref{fig:3}(c2)) in a mechanism that closely resembled a skewering mechanism proposed earlier~\cite{https://doi.org/10.1002/fld.803, KAWAGUCHI1997325}. We note that such skewering propagation was reported in miscible and immiscible displacement experiments involving non-Newtonian fluids and in numerical simulations of miscible viscous fingering involving a Newtonian fluid pair of high mobility ratio~\cite{KAWAGUCHI1997325, https://doi.org/10.1002/fld.803}.
Eventually, we noted that multiple new fingers grew at the broadened fingertip. These new fingers propagated in a non-linear fashion, while interacting with one another to form trailing lobes. We identified this mechanism as dense branching (DB). The skewering mechanism is seen to coexist with another novel finger propagation mechanism, wherein a few fingers propagate orthogonally to the flow direction for a small period of time (indicated by pink dotted line in Figs.~\ref{fig:3}(c1-c4)). We note that the zigzag and skewering propagation mechanisms reported here occurred during the entire duration of pattern growth. 

A freshly prepared clay is composed not of individual clay particles, but of 1-D stacks called tactoids having a wide distribution of sizes~\cite{doi:10.1021/la402478h}. Some of these tactoids remain unbroken even as the sample becomes highly elastic because of physical aging~\cite{ALI201585}. When the suspension elasticity is low at an earlier stage of aging, the suspension medium is composed of disconnected strands of self-assembled clay tactoids and exfoliated particles~\cite{D1SM00987G}. For finger propagation that is dominated by viscous forces, the presence of unbroken tactoids and disconnected strands presents a heterogeneous environment to the growing finger. The resultant stochastic variations in permeability lead to the emergence of preferential flow pathways that allow the advancing fingers to propagate with high mobility. We conclude therefore that channeling flow due to the heterogeneous permeability of the displaced clay drives the observed snake-like zig-zag finger propagation profile, finger growth \textit{via} skewering and the orthogonal movement of fingers.

We next analysed finger propagation mechanisms at the clay–water interface (pure clay or clay with DMF and TSPP in Fig.~\ref{fig:3}) by computing some standard metrics. Finger coalescence events were quantified by estimating the total number of trailing lobes and their average areas as a function of elapsed time since the start of the experiment, \textit{i.e.}, for the entire duration after the displacing fluid was injected into the Hele-Shaw cell. This data is plotted in Figs.~\ref{fig:5}(a,b). We see from Fig.~\ref{fig:5}(a) that the number of trailing lobes grows monotonically with time and is highest for the clay with added TSPP. We attribute this to frequent mergers of the snake-like zig-zag finger propagation profiles.  Simultaneously, the average areas of the trailing lobes, plotted in Fig.~\ref{fig:5}(b), are seen to be highest for clay with added DMF due to the skewering of fingertips before they split and grew. 
 
 \subsection{Finger propagation profiles during miscible displacements of aqueous clay samples prepared with NaCl, KCl and glucose}
\begin{figure}[!t]
		\includegraphics[width=0.49\textwidth]{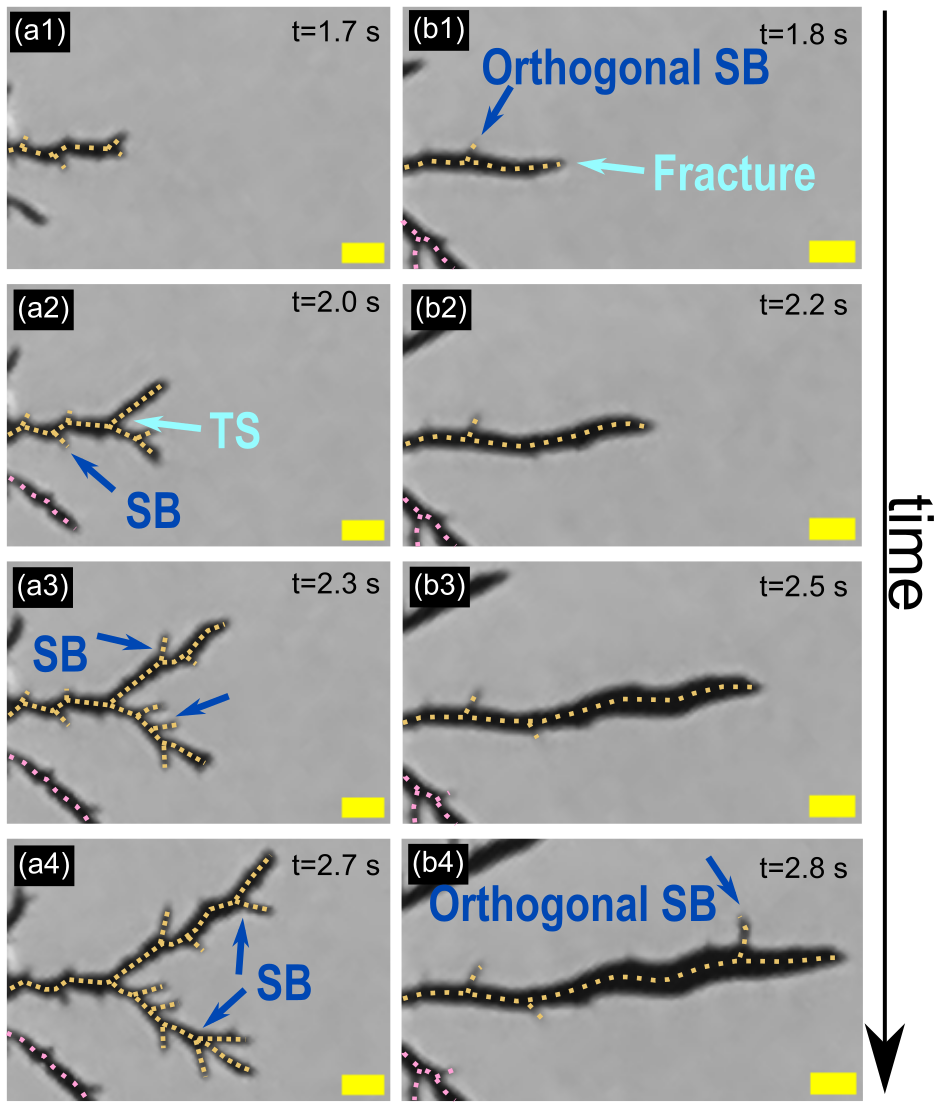}
		\centering
		\caption{\label{fig:4} Cropped grayscale images displaying the temporal propagation profiles of fingers formed during the miscible displacement of aged clay of concentration of 3\% w/v with  \textbf{(a1-a4)} 1 mM NaCl and \textbf{(b1-b4)} 2 mM NaCl added to the suspension medium. Dotted lines have been drawn for clarity. Each scale bar represents 0.5 cm. The abbreviations used are as follows: TS: Tip-splitting and SB: Side branching. Interfacial patterns due to the displacement of highly rigid clays did not form trailing lobes and are therefore excluded from the trailing lobes analysis displayed earlier.
  }
  \end{figure}

The presence of NaCl, KCl and glucose in the aqueous medium enhances the elasticity of the clay suspension by accelerating its physical aging in a process that is driven by the accelerated self-assembly of clay particles. 
Figures~\ref{fig:4}(a1-a4,b1-b4) display the evolution of interfacial fingers when highly elastic clays, prepared by adding NaCl of concentrations of $C_A = 1$ mM and $C_A=2$ mM to the suspension medium, respectively, were displaced by injecting water in the Hele-Shaw cell. The movies from which these snapshots were extracted are provided as Supplementary Movies~4 and 5, respectively. For these displacement experiments, the fluid-suspension interface is comparatively sharp, the spacings between the neighboring fingers increase significantly and no coalescence events were noted. 

For the clay samples with $C_A$ = 1 mM NaCl, we see from Figs.~\ref{fig:4}(a1-a4) that finger propagation is dominated by side branching, though occasional tip-splitting events are also noted. Side branching, which refers to the development of smaller, secondary fingers behind the tip of the main finger, was reported when a shear-thinning fluid was displaced by a Newtonian fluid~\cite{PhysRevLett.80.1433}. There is significant viscosity-thinning at the propagating tip of the advancing finger due to the large applied shear. Simultaneously, the finger broadens due to the applied injection pressure. The competition between these two factors results in the observed shedding of side branches (denoted as SB in Fig.~\ref{fig:4}). In agreement with previous reports~\cite{liquid-crystal,doi:10.1126/science.243.4895.1150,PhysRevLett.80.1433}, we note that displacement of the shear-thinning clay samples results in large flow anisotropy, with the tip experiencing lower viscosity along the flow direction compared to the lateral direction. This affects the growth of the invading water fingertip and forms multiple side-branches as fingers propagate. We therefore conclude that when the displaced clay was characterized by a moderately high plateau elastic moduli ranging between $800 \leq {G_L}^\prime \leq 950$ Pa (Fig.~\ref{fig:2}(c)), achieved by the introduction of additives such as NaCl, KCl and glucose in the suspension medium to enhance clay aging, pattern growth was dominated by significant side-branching and occasional tip-splitting events.

  When a clay sample containing the highest concentration of polar salt, 2 mM NaCl, was displaced by dyed water (Figs.~\ref{fig:4}(b1-b4)), we observed comparatively thicker interfacial fingers with pointed tips. These fingers display side branching with branching angles close to 90$^\circ$, as marked in Fig.~\ref{fig:4}(b1,b4). Such orthogonal branching, a typical characteristic of fracture propagation, is noted by us only in those displacement experiments involving clay samples of elastic modulus, $G_L^\prime > 10^3$ Pa. 
  In miscible displacement experiments such as the ones reported here, large elastic stresses are expected to develop below the clay suspension yielding point when the sample contains very high amounts of polar additives like NaCl and KCl. The observed fractures emerge due to the rapid release of these elastic stresses if the energy required to create two new surfaces is less than that required for system-wide yielding. Since the linear elastic moduli of samples with glucose are much smaller in comparison to clays with added polar salts (Fig.~\ref{fig:2}(c)), we did not observe any fracture patterns in these samples (Supplementary Information Fig.~S4). We note that it may still be possible to generate fracture patterns in the presence of glucose by allowing the clay to age for a longer time or by enhancing the glucose content in the suspension medium. An inspection of Fig. S4(b), which displays the interfacial pattern generated during the displacement of a clay sample with a large glucose content, reveals an instability that closely resembles a fracture pattern.

   \begin{figure}[!t]
		\includegraphics[width=0.99\textwidth]{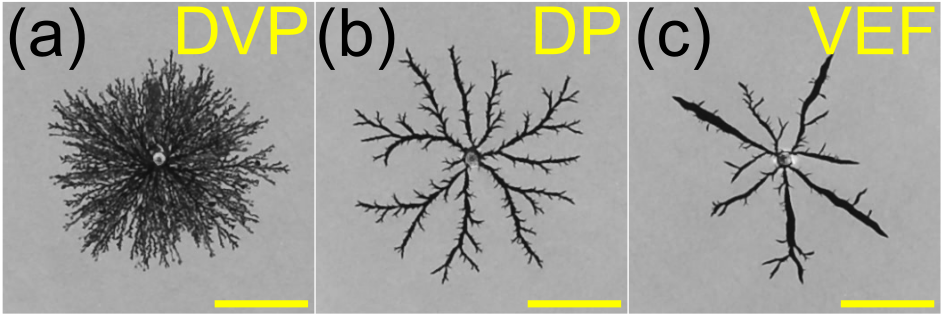}
		\centering
		\caption{\label{fig:6} Representative grayscale images displaying (a) dense viscous pattern (DVP), (b) dendritic pattern (DP) and (c) viscoelastic fractures (VEF) observed during the miscible displacements of clay at an aging time of \( t_w = 5 \) h. The clay samples in these images were captured under three conditions: without any additive, with 1 mM NaCl, and with 2 mM NaCl, respectively. The scale bar{s are} 5 cm.
  }
  \end{figure}
  \subsection{Classification of fully developed interfacial pattern morphologies}
  We identified three distinct interfacial patterns, \textit{viz.} dense viscous patterns, dendritic patterns and viscoelastic fractures, by inspecting the interfacial patterns displayed in Figs.~S3 and S4. Figure~\ref{fig:6}(a) shows a representative image of a dense viscous pattern (DVP). DVPs were generated during the displacement of pure clay samples prepared without any additive or in samples containing chaotropic additives like DMF and TSPP. These compact interfacial patterns were characterized by tip-splitting, channeling and frequent coalescence events. DVP growth is predominantly driven by viscously unstable flow in homogeneous or heterogeneous porous media. These patterns were therefore seen only when the displaced clay had lower rigidity, $G_L^\prime \leq 573$ Pa, and did not display strong nonlinear rheology. Dendritic patterns (DP, Fig.~\ref{fig:6}(b)) were characterized by dominant side-branching and were formed when clay samples of comparatively larger rigidities, achieved by adding moderate concentrations of additives like polar salts (KCl and NaCl) and glucose, were displaced. In contrast to DVP, DP growth is expected to be governed by the shear-thinning rheology of the displaced clay. Viscoelastic fractures (VEF, Fig.~\ref{fig:6}(c)) were observed for the most elastic clay samples ($G_L^\prime \geq 1089$ Pa), prepared by increasing the concentrations of polar salts and glucose in the suspension medium. These interfacial instabilities closely resemble the fracture patterns observed in brittle materials and are characterised by a very small number of primary fingers, fast fingertip propagation and the frequent shedding of side branches in a direction orthogonal to the primary fingers.

\subsection{Characterisation of fully developed interfacial pattern morphologies}
  \begin{figure}[!t]
		\includegraphics[width=0.6\textwidth]{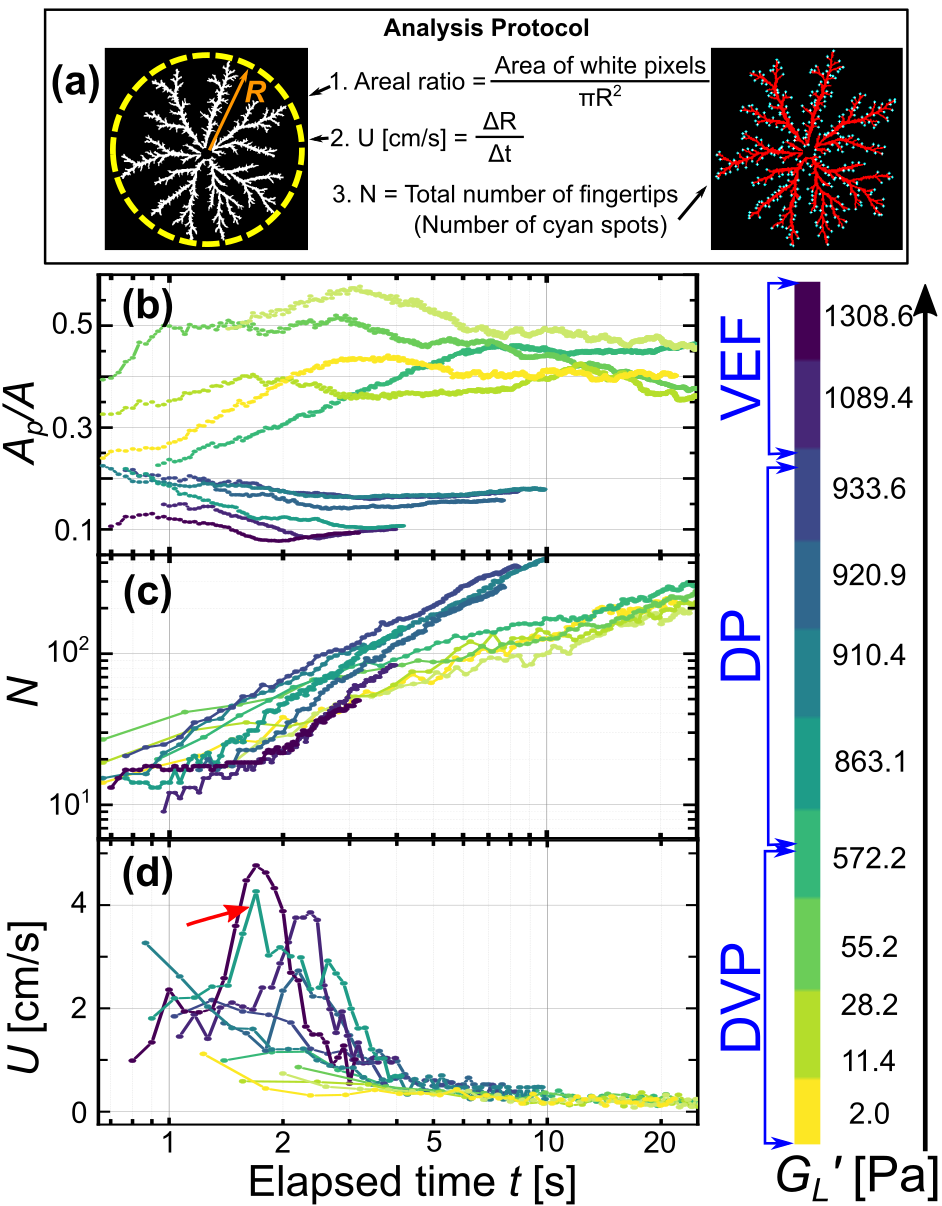}
		\centering
		\caption{\label{fig:7} (a) Schematic of analysis protocol for estimating areal ratio, $A_P/A$, instantaneous fingertip velocity of the longest propagating finger, $U$ and total number of fingertips, $N$.  (b) $Ap/A$, (c) ${U}$ and (d) $N$ as a function of elapsed time, $t$. The scale bar on the right maps the values of $G_L^\prime$ to the observed interfacial patterns.
  }
  \end{figure}
We have demonstrated here that the growth kinetics and final morphologies of interfacial patterns, formed during the displacement of clay by water in a radial Hele-Shaw geometry, are sensitive to the plateau elastic modulus $G_L^\prime$ of the displaced clay. We next characterised the patterns by estimating certain metrics, the protocols for which are schematically illustrated in Fig.~\ref{fig:7}(a) and discussed in more detail in Supplementary Information Section~ST2. Figs.~\ref{fig:7}(b-d) display our analyses of all the observed pattern morphologies using the three independent metrics. Fig.~\ref{fig:7}(b) shows the temporal evolution of the pattern areal ratio{s}, $A_p/A$, for all the patterns. Here, $A_p$ is the areal coverage of the interfacial pattern and $A$ is the area of the smallest circle enclosing the entire pattern at that instant. This parameter was proposed in earlier work~\cite{PALAK2022100047,PALAK2023100084} to quantify the extent of the spatial coverage of the interfacial pattern. Fig.~\ref{fig:7}(c) plots the total number of fingertips, $N$, for all the patterns studied here, and indicates the combined effects of tip-splitting, side branching and coalescence. Finally, Fig.~\ref{fig:7}(d) shows the instantaneous fingertip velocity ${U}$ of the longest propagating finger, which parametrizes the speed of the interfacial dynamics. The scale bar on the right maps the values of $G_L^\prime$ to the observed interfacial patterns. 
 
The plots in Figs.~\ref{fig:7}(b-d) reveal that the observed DVP, DP and VEF patterns display unique features. From our analyses of $A_p/A$ in Fig.~\ref{fig:7}(b), we see that dense viscous patterns (DVPs) are the most compact/ space-filling, while the total number of fingertips, $N$, increases monotonically with time for all the patterns, as seen from Fig.~\ref{fig:7}(c). Finally, we see from Fig.~\ref{fig:7}(d) that the fingers in DVP propagate at the slowest speeds, with the propagation velocity $U$ reaching a steady state over time. These features of DVPs result from the interplay of several propagation mechanisms including tip-splitting, shielding, coalescence and zig-zag propagation as described earlier (Figs.~\ref{fig:3}(a1-c4)). Dendritic patterns (DP), in contrast, are relatively less compact, as indicated by the intermediate values of $A_p/A$ ratios plotted in Fig.\ref{fig:7}(b). We further note that the number of fingertips, $N$, shown in Fig.~\ref{fig:7}(c), increases more rapidly for DP than for DVP due to the rapid formation of side branches, an ubiquitous characteristic of dendritic growth. The propagation velocity $U$ for DP, shown in Fig.~\ref{fig:7}(d), is faster than that observed for DVP at initial times and eventually slows down with time due to the shedding of multiple side branches. Viscoelastic fractures (VEF) are the least compact of the three morphologies, as indicated by their lowest $A_p/A$ values in Fig.~\ref{fig:7}(b). Since tip-splitting and side branching are minimal during the growth of VEF patterns, their morphologies exhibit a very small number of fingertips, $N$ (Fig.~\ref{fig:7}(c)). Lastly, we note that VEF pattern growth shows rapid finger propagation during the initial stages, with pattern growth seen to be substantially inhibited at later times due to orthogonal side-branch formation. The propagation velocity of the longest finger in a VEF therefore shows a non-monotonic propagation velocity, $U$. Interestingly, one of the dendritic patterns studied by us, marked by a red arrow in Fig.~\ref{fig:7}(d), exhibits a non-monotonic velocity profile, suggesting that this pattern lies in a crossover zone between DP and VEF morphologies.

\subsection{Equivalence between increasing clay concentration and incorporation of additives while tuning interfacial pattern selection}
We next performed displacement experiments using pure clay samples of increasing concentrations and observed a transition from DVP to DP. The interfacial pattern morphologies are presented in Section~ST3 of the supplementary information. We had shown in an earlier work~\cite{PALAK2022100047,PALAK2023100084} that a transition from DVP to DP and eventually to VEF can be achieved in Hele-Shaw displacement experiments involving pure clay and water. This transition was achieved by increasing the age of the displaced clay or by reducing the injection flow rate of the displacing water. In the present work, we have demonstrated two additional protocols by which clay displacement patterns can be fine-tuned. While adding appropriate quantities of dissociative salts like NaCl, KCl and TSPP serves as an effective strategy to control the interparticle attraction strength between clay particles, the incorporation of non-ionic additives like sucrose and DMF can alter clay elasticity by tuning the hydrogen bonding in the underlying suspension medium. Even though the mechanisms controlling the rates of clay aging in the presence of these different additives (\textit{i.e.}, increased charge screening in the presence of NaCl and KCl, electrostatic repulsion in the presence of TSPP and enhancement and inhibition of hydrogen bonding in the presence of glucose and DMF respectively) are very different, we have demonstrated that interfacial morphologies such as DVP, DP and VEF can be easily generated during miscible displacements of aqueous clay by tuning additive nature and content. This is feasible since interfacial pattern growth is dictated solely by the bulk linear elasticity of the displaced clay. 

A previous study by Saha et al.~\cite{doi:10.1021/acs.langmuir.5b00291} presented a comprehensive superposition of the relaxation dynamics of Laponite clay when clay and NaCl concentrations and medium temperature were increased in separate experiments. It was argued that enhancing all these parameters resulted in the accelerated dissolution of $Na^+$ in the suspension medium, which resulted in self-similar energy landscapes of the soft glassy clay samples. We show here that the same sequence of interfacial displacement patterns may be generated as in previous experiments~\cite{PALAK2022100047,PALAK2023100084} by changing an entirely different set of physicochemical parameters, \textit{viz.}, the chaotropic or kosmotropic salt content in the aqueous suspension medium, or by appropriately tuning the concentration of the displaced clay. The present work identifies that fine-tuning clay concentration, age, displacing fluid injection rate, and the content and nature of the incorporated additives have equivalent effects on the global characteristics of the interfacial patterns formed during confined fluid displacements.

\section{Conclusions}
  We have demonstrated that the linear elastic modulus, determined from the plateau value of the elastic modulus, decides the morphologies and growth kinetics of the interfacial patterns that are generated during the confined displacements of clay samples by miscible water in a radial Hele-Shaw geometry. The addition of salts can change the aging of clay \textit{via} the following 2 mechanisms: it either tunes the interparticle electrostatic interaction or changes the structure of the aqueous suspension medium. Regardless of the mechanism, adding salts allows us to sensitively control the aging dynamics and hence the evolution of the bulk elastic modulus of the displaced clay suspension~\cite{D1SM00987G}. We hypothesized, therefore, that controlling clay aging \textit{via} the controlled incorporation of different additives should serve as an easy and practical method for programming the morphologies and growth kinetics of interfacial patterns during the confined miscible displacements of aging aqueous colloidal Laponite clay.

  Our study also reveals novel local mechanisms of pattern growth, such as zig-zag propagation and skewering of the dense viscous patterns. We believe that these novel propagation mechanisms arise from the coupling of interfacial viscous instabilities with heterogeneities in the permeability of the displaced clay. While zig-zag finger propagation was predicted theoretically~\cite{10.1063/1.475259}, this unique mechanism has never been demonstrated experimentally in the available scientific literature. Skewering patterns have been previously reported during the displacement of Newtonian and polymeric fluids~\cite{https://doi.org/10.1002/fld.803, KAWAGUCHI1997325}. We demonstrated here that skewers (Fig.~\ref{fig:3}(c)) can also form due to the presence of stochastic heterogeneity in the permeability of the displaced aqueous colloidal clay. Since the clay samples did not age substantially during the displacement experiments reported here, the stochastic permeability heterogeneity profile may be assumed to be stationary as water invades clay. We note that if the injection speed of the displacing water is reduced substantially, such that the clay sample ages during the displacement experiment, an even wider range of novel pattern morphologies and pattern growth mechanisms can be obtained.
  

  By incorporating chaotropic and kosmotropic additives to the suspension medium while preparing aqueous clay samples, we achieved smooth transitions between a range of interfacial patterns, \textit{viz.} dense viscous patterns, dendritic patterns and viscoelastic fractures. 
  We demonstrated that modifications in inter-Laponite interactions or disruption of the hydrogen bonds in the aqueous suspension medium can both play crucial roles in determining the local finger propagation profile when the elastic modulus of the displaced clay is small. We analyzed all the patterns observed in our experiments using three key metrics: (i) the areal ratio \( A_p/A \), which indicates the spatial coverage of the pattern (ii) fingertip velocity \( U \), which reflects the speed of pattern growth and (iii) the number of fingertips \( N \), which summarizes the combined effects of tip-splitting, side-branching and coalescence. We showed that these metrics have unique signatures for the three independent pattern morphologies observed in our experiments.

Interfacial pattern morphologies formed by displacing aqueous clay suspensions in confined geometries can therefore be fine-tuned by incorporating additives that enhance or delay the physical aging behavior of the samples, thereby increasing or decreasing their linear elastic moduli. This strategy can also be exploited to significantly enhance or completely suppress the growth of interfacial patterns. In future experiments, it would be interesting to control the heterogeneous permeability resulting from the unbroken clay tactoids and disconnected strands of clay particles. The development of clay elasticity and permeability heterogeneity on the one hand, and the finger propagation profiles during displacement experiments on the other, must be correlated. It would be interesting to study these correlations in systematic experiments. It also follows that an even larger array of interfacial patterns can be generated simply by tuning the miscibilities of the fluid pairs. We conclude by noting that the potential to minutely control interfacial displacement patterns can also have far-reaching implications in materials processing applications. The suppression of instabilities during the transport of clay slurries, for example, can maximize transport efficiency while reducing energy costs. 
%
%

\ack{We would like to thank Dr. Palak for co-developing the codes for pattern analysis.}


\roles{ \textbf{VRS Parmar}: Investigation (lead), Formal analysis (lead), Writing– original draft (lead), Methodology (equal). \textbf{RB}: Conceptualization (lead), Resources (lead), Supervision (lead), Writing– review \& editing (equal).}

\data{Data will be made available on request.}

\suppdata{\textbf{Movie 1.} (1 MB mp4) Fingertip splitting events when a clay sample of concentration, $C_L=$ 3\% w/v, was displaced by water.
\textbf{Movie 2.} (4.6 MB mp4) Snakelike zig-zag finger propagation profile when a clay sample of concentration, $C_L=$ 3\% w/v, and additive concentration 2 mM TSPP was displaced by water.
\textbf{Movie 3.} (2.4 MB mp4) Skewering and dense branching events when a clay sample of concentration, $C_L=$ 3\% w/v, and additive concentration 260 mM DMF was displaced by water.
\textbf{Movie 4.} (0.4 MB mp4) Frequent side-branching and occasional tip-splitting events when a clay sample of concentration, $C_L=$ 3\% w/v, and additive concentration 1 mM NaCl was displaced by water.
\textbf{Movie 5.} (0.4 MB mp4) Fracture propagation and orthogonal side branching events when a clay sample of concentration, $C_L=$ 3\% w/v, and additive concentration 2 mM NaCl was displaced by water.
\textbf{Supplementary material.} (PDF) Viscous modulus of clay samples and details of analysis protocols used in this study. In addition, all the global pattern morphologies were obtained in this study for various physicochemical conditions of clay.}

\bibliographystyle{vancouver}
\bibliography{ref2}


\section*{Supplementary material}

# Supplementary Material

# Permeability heterogeneity and bulk linear elasticity of displaced clay suspensions determine interfacial pattern morphologies in Hele-Shaw experiments

Vaibhav Raj Singh Parmar[1,†] and Ranjini Bandyopadhyay[1,*]

[1]*Soft Condensed Matter Group, Raman Research Institute, C. V. Raman Avenue, Sadashivanagar, Bangalore 560 080, INDIA*

September 6, 2025

[†]vaibhav@rri.res.in
[*]Corresponding Author: Ranjini Bandyopadhyay; Email: ranjini@rri.res.in

1

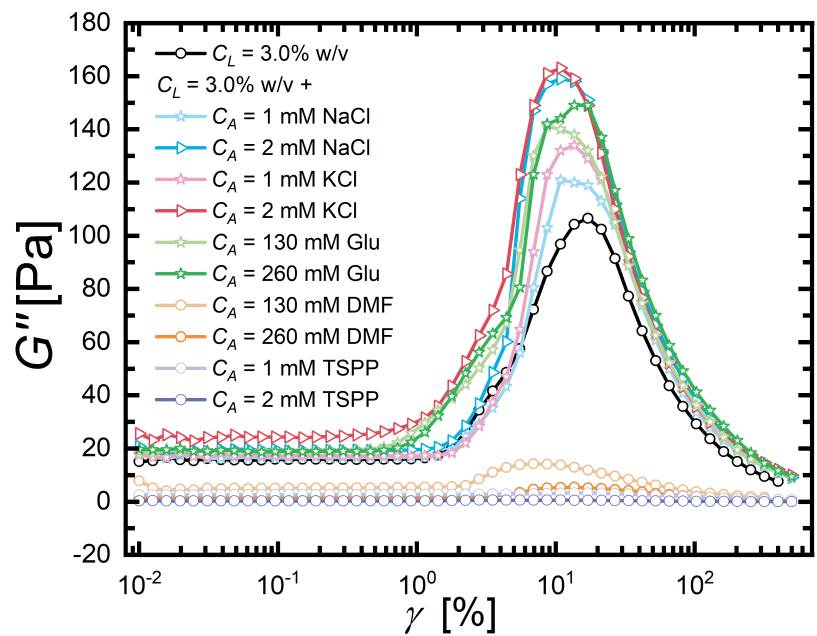

Supplementary Fig. S1: Semi-log plots of amplitude sweep oscillatory rheological measurements showing viscous moduli $G''$ for 3.0% w/v clay suspensions with and without additives, measured at $t_w = 5$ h.

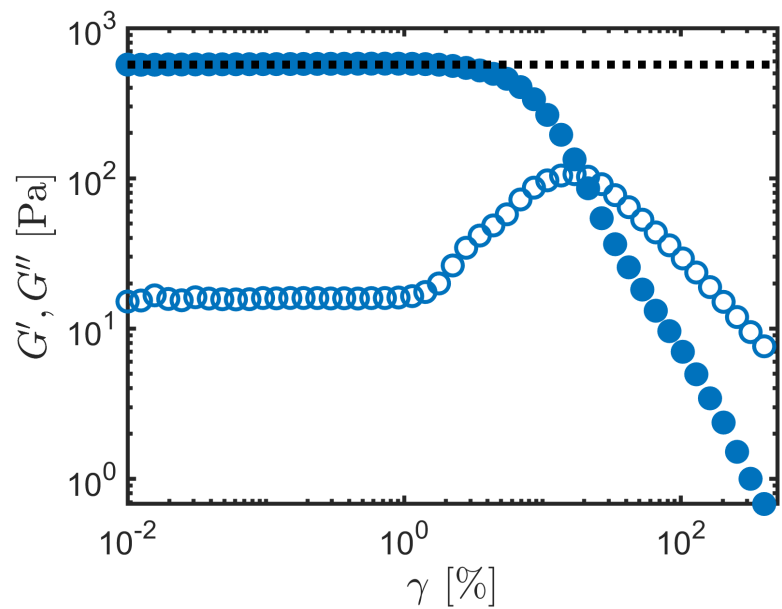

Supplementary Fig. S2: Elastic ($G'$, solid symbols) and viscous ($G''$, hollow symbols) moduli of a clay sample of concentration 3.0% w/v without any additive at $t_w = 5$h. The linear elastic modulus, $G_L'$, is equal to the magnitude of $G'$ at a very low applied strain amplitude, $\gamma$. $G_L'$ was extracted by fitting a straight horizontal line (black dotted line) to $G'$ at $\gamma \leq 0.578$.

3

# ST1 Interfacial pattern morphologies during displacement of clay in the presence of additives that delay or accelerate aging

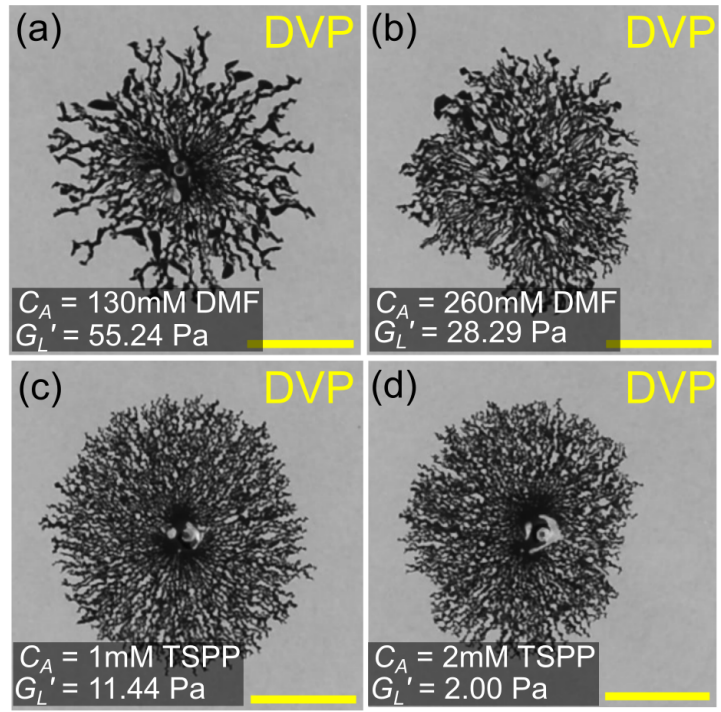

Supplementary Fig. S3: Grayscale images of interfacial patterns formed when aqueous clay samples of concentration 3.0% w/v and aging time $t_w = 5$h in the presence of DMF and TSPP were displaced by dyed water. Images are presented for clay samples with (a) 130 mM dimethylformamide (DMF), (b) 260 mM DMF, (c) 1 mM tetrasodium pyrophosphate (TSPP) and (d) 2 mM TSPP. The displacing fluid (dyed water) is injected at a fixed injection flow rate $q = 5$ ml/min. The linear elastic moduli, $G'_L$, of clay samples are indicated at the bottom left. The scale bars are 5 cm.

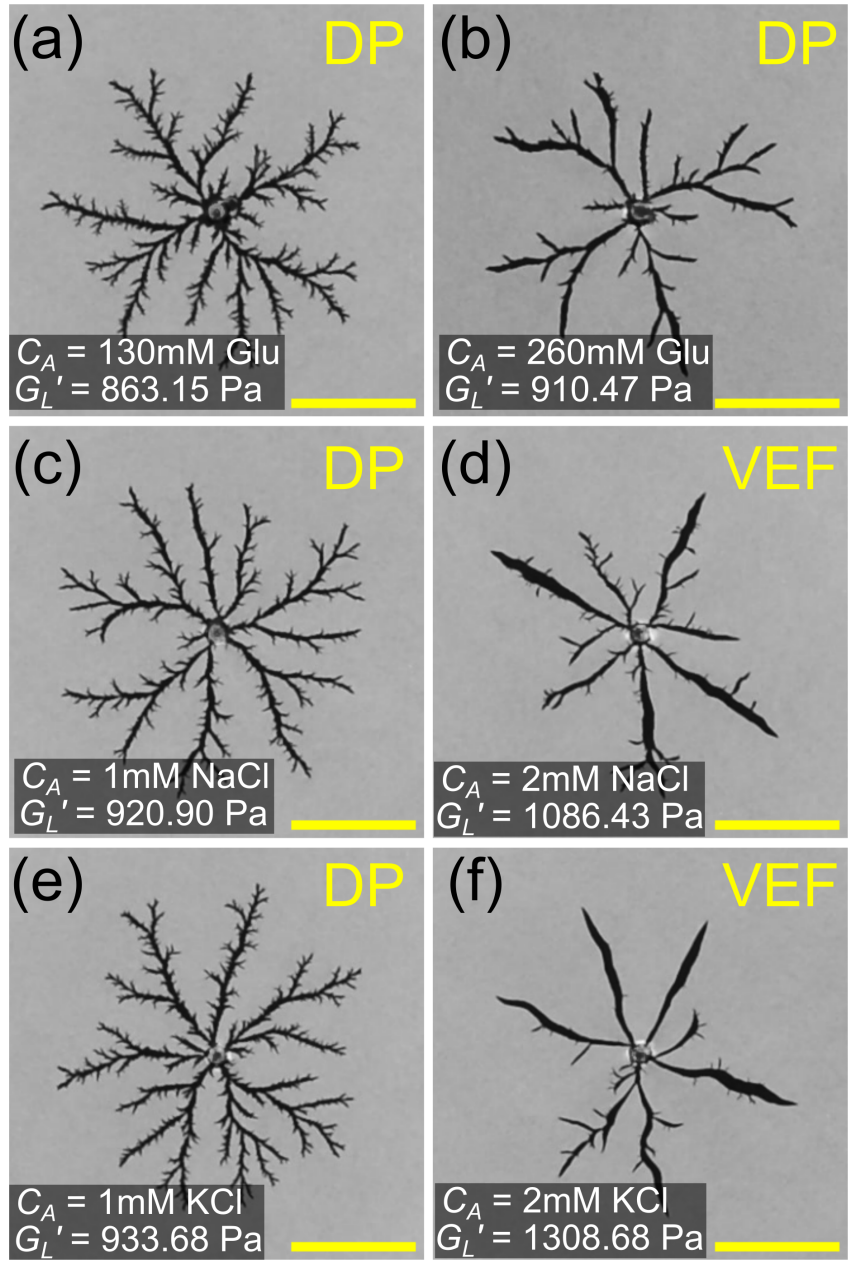

Supplementary Fig. S4: Grayscale images of interfacial patterns during the displacement of aqueous clay of concentration 3.0% w/v and aging time $t_w = 5$h in the presence of (a) 130 mM glucose, (b) 260 mM glucose, (c) 1 mM NaCl, (d) 2 mM NaCl, (e) 1 mM KCl, and (f) 2 mM KCl. The displacing fluid (dyed water) is injected at a fixed injection flow rate $q = 5$ ml/min. The linear elastic moduli, $G'_L$, of clay samples are indicated at the bottom left. The scale bars are 5 cm.

5

# ST2 Calculation of areal ratio $A_p/A$, instantaneous fingertip velocity $U$ and total number of fingertips $N$

The areal ratio, $A_p/A$, is defined as the ratio of the area of the interfacial pattern to the area of the smallest circle that encloses the entire pattern. As shown in Fig. 8(a) of the main manuscript, we calculated $A_p$ by estimating the number of white pixels in the binarised image of the pattern and multiplied it by the area of one pixel (pixel area $\sim 0.2$ mm$^2$). Next, we divided this number by $A = \pi R^2$, where $R$ is the length of the longest finger and defines the radius of the smallest circle enclosing the entire pattern. We calculated the instantaneous fingertip velocity by tracking the tip of the longest propagating finger and estimating the first-order derivative $U = \frac{\Delta R}{\Delta t}$ using the `diff` function in MATLAB. Finally, we estimated the number of fingertips, $N$, by making a skeleton of the pattern using the `bwmorph` function in MATLAB. These skeletons are 1-pixel wide structures. For each snapshot, we identified the fingertips by locating the white pixels in the skeleton with exactly one neighboring white pixel (endpoints of skeleton structures). This method accurately counts the fingertips formed by both tip-splitting and side-branching mechanisms during the growth of an interfacial pattern.

# ST3  Transitions between interfacial pattern morphologies with increasing clay concentration

Figures S5(a-c) display the interfacial patterns formed when dyed water radially displaces aqueous clay of increasing concentrations at a fixed sample age, $t_w = 5$ h. For a low clay concentration $C_L = 3.0$ % w/v, a dense viscous pattern, with interfacial growth dominated by finger-tip splitting events and nominal side branching, was observed (Fig. S5(a)). As clay concentration was increased (Figs. S5(b,c)), we observed that the fluid-clay interface became increasingly sharp, and pattern growth was dominated by side-branching events. We did not observe viscoelastic fractures in this set of experiments due to the weak elastic response of the displaced clay in the narrow concentration range used here. However, we know from a previous study [1] that fractures can be generated during the miscible displacement of pure clay samples whose elasticities were substantially higher due to aging over longer durations ($\sim$ 24h).

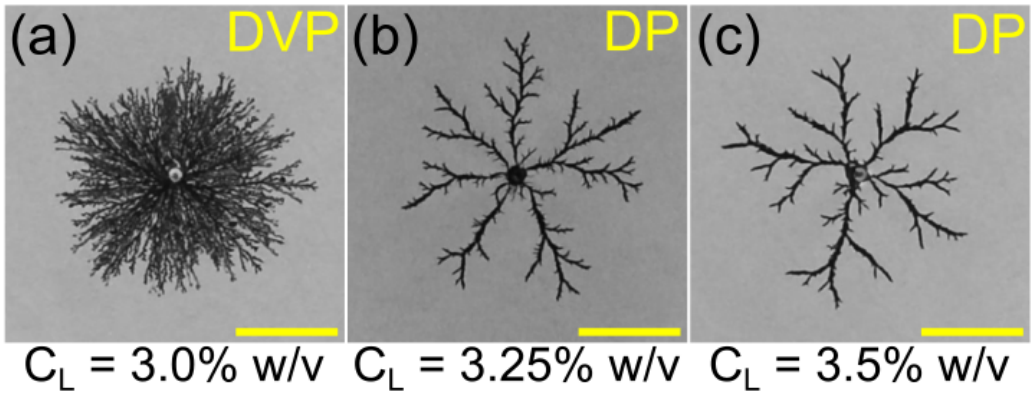

Supplementary Fig. S5: Grayscale images of interfacial pattern morphologies formed when aqueous clay of age $t_w = 5$ h and concentrations (a) 3.00% w/v (b) 3.25% w/v and (c) 3.50% w/v were displaced by dyed water at a fixed injection flow rate $q = 5$ ml/min. The scale bar is 5 cm.

# References

[1] Palak, Vaibhav Raj Singh Parmar, Debasish Saha, and Ranjini Bandyopadhyay. Pattern selection in radial displacements of a confined aging viscoelastic fluid. *JCIS Open*, 6:100047, 2022.


\end{document}